\documentstyle[preprint,aps]{revtex}

\begin{document}

\draft         


\title{ Quantum evolution of inhomogeneities \\
in curved space}

\author{H.\ C.\ Reis \cite{hcr} }
 
\address{ Departamento de Raios C\'{o}smicos e Cronologia \\
 Universidade Estadual de Campinas - Unicamp \\
 CEP: 13083-970, Campinas - SP, Brazil }

\maketitle

\begin{abstract}

        We obtain the renormalized equations of motion for matter and
        semi-classical gravity in an inhomogeneous space-time.  We use
        the functional Schr\"odinger picture and a simple Gaussian
        approximation to analyze the time evolution of the $\lambda
        \phi^4$ model, and we establish the renormalizability of this
        non-perturbative approximation. We also show that the
        energy-momentum tensor in this approximation is finite once we
        consider the usual mass and coupling constant renormalizations, 
        without the need of further geometrical counter-terms.

\end{abstract}

\vskip 1.in

\begin{center}
{\em Accepted to publish in Int. J. Mod. Phys.}{\bf A}
\end{center}

\newpage



\section{Introduction}      
\label{sec:int}


One of most rich and interesting subject in physics is the early
universe, because it brings together two other topics that apparently
have no connection: elementary particle physics and gravitation. 
According the standard cosmological model ({\em Big Bang}), 
the early universe was in such a high temperature state that
particle physics must be used to describe the matter in this
condition.  The meaning of this is that the evolution of universe in
the initial instants depended upon the particle spectrum and their
interactions. In particular, the dynamics of scalar fields could give
rise to the appearance of an inflationary era$^{\cite {abb}}$, that
occurred during and/or after one of many phase transitions that the
universe experienced.

In general, the study of inflationary scenarios is made considering
only some aspects of scalar field dynamics responsible for the
existence of an inflationary phase. For instance, there are analyses
of the classical effects of inhomogeneities in a scalar field in
homogeneous spaces$^{\cite{alb}}$, as well as in inhomogeneous 
spaces$^{\cite{suo}}$. On the other hand, the studies of quantum and thermal 
effects do not that into account the consequences of departures from
homogeneous situations$^{\cite{ebo,hug}}$.

The purpose of this paper is to obtain the equations of motion for the
$\lambda \phi^4$ model and semi-classical gravity, taking into account
non-perturbative quantum effects and inhomogeneities. This work
is a generalization of ref.{\cite{hug}} which analyzed the time
evolution of homogeneous scalar fields in a Robertson-Walker space-time.
This was accomplished in this reference using the functional
Schr\"odinger picture and a simple Gaussian approximation. In this case
it was possible to show the renormalizabi\-lity of the equation of motion
and energy-momentum tensor. In order to carry out a consistent analysis
of the time evolution of inhomogeneities we must consider inhomogeneous
spaces, however, we assumed that space is inhomogeneous in a single
direction$^{\cite{suo}}$ to the problem be tractable.  Furthermore, the
exact solution of time evolution of an interacting system is impossible
to be obtained, except for system whose Hamiltonian is quadratic, and
consequently, approximation methods are needed. Therefore, we use
variational approximations$^{\cite{jac,ker,bal}}$ which lead to
tractable equations to the various parameters defining a Gaussian {\em
Ansatz}. Although we work with a restrict class of states, the use of
this variational methods allow us to obtain some non-perturbative
effects through self-consistent equations.

The inclusion of quantum effects in the evolution of the system gives
rise to divergences in the equations of motion and the energy-momentum
tensor. Therefore, we obtain the renormalization prescription for
the mass and coupling constants in order to remove the infinities from
the equations of motion. We also show that this renormalization
prescription is enough to render the energy-momentum tensor finite in
our approximation without the introduction of further geometrical
counter-terms.  In brief, our procedure leads to a consistent set of
finite equations to matter and semi-classical gravitation in the
presence of inhomogeneities.

In this work we make use of the functional Schr\"odinger
picture$^{\cite{sch}}$ to set up the equations of motion for this
initial value problem. The conventional formulation of quantum field
theory in terms of causal Green's functions in the Heisenberg picture
is not specially suited to time-dependent problems that employ an
initial condition for specific solution. Green's functions contain all
information needed to determining transition rates, $S$-matrix
elements, etc., of system in equilibrium where initial data are
superfluous. However, following a system's time evolution from a
definite initial configuration is more efficiently accomplished in the
Schr\"odinger picture description, where the initial data consist of
specifying a state.

For bosonic fields, the functional Schr\"odinger picture is a
generalization from ordinary quantum mechanics to the infinite number
of degrees of freedom that constitute a field. Therefore, we can use
the mathematical/physical intuition acquired in quantum mechanics to
analyze field-theoretic problems. However, the functional
Schr\"odinger picture is not as widely used in actual calculations as
the Green's function method since renormalization is more easily
carried out in the latter framework.  Nevertheless, it has been
established renormalizability to the Schr\"odinger picture for
both static$^{\cite{sym}}$ and time-dependent$^{\cite{sam}}$ cases. In this
work we expand the list of situations where the renormalizability of
this framework can be shown.

This work is organized as follows: In Section \ref{gen} we present the
functional Schr\"odinger picture for a scalar field, as well as the
variational principle used to obtain the equations of motion of the
system. Section \ref{model} contains the $\lambda \phi^4$ model in which
the homogeneous hypothesis in one direction ($z$ direction) is
abandoned and also our simplified Gaussian {\em Ansatz}. In section 
IV we renormalize the effective action in the Centrella-Wilson metric, 
which leads to the renormalized equations of motion for the variational 
parameters. The analysis of the renormalization of the energy-momentum 
tensor is shown in Sec. V. We draw our conclusions in Sec.VI.


\section{Functional Schr\"odinger picture}
\label{gen}


In the Schr\"odinger picture$^{\cite{sch}}$, scalar field states are
described by wave functionals $\Psi(\phi)$ of a $c$-number $\phi({\bf
x})$ at a fixed time,

\begin{equation}
\Psi(\phi) = \langle \phi | \Psi \rangle \; . 
\end{equation}
The inner product is defined by functional integration,

\begin{equation}
\langle \Psi_1 | \Psi_2 \rangle \Leftrightarrow \int D\phi~ \Psi_1^*
(\phi) \Psi_2(\phi) \; ,
\end{equation}
while operators are represented by functional kernels $O(\phi,
\phi^\prime)$,

\begin{equation}
O | \Psi \rangle \Leftrightarrow \int D\phi^\prime~ O(\phi, 
\phi^\prime) \Psi(\phi^\prime) \; .
\end{equation}

For the canonical field operator at a fixed time $\phi({\bf x})$ (the
time argument is common to all operators in the Scr\"odinger picture,
so it is suppressed), we adopt a diagonal kernel $\Phi({\bf x})
\Rightarrow \phi({\bf x)} \delta(\phi - \phi^\prime)$; the canonical 
commutation relations determine the canonical momentum kernel to be 
$\Pi({\bf x}) \Rightarrow (1/i) [\delta/\delta\phi({\bf x})] 
\delta(\phi - \phi^\prime)$. In this way $\Phi$ acts by multiplication on 
functionals of $\phi$ and $\Pi$ acts by functional differentiation.
Hence, the action of any operator constructed from $\Pi$ and $\Phi$ is

\begin{equation}
{\cal O}(\Pi, \Phi) | \Psi \rangle \Rightarrow 
{\cal O}\left [ \frac{1}{i} \frac{\delta}{\delta \phi}, \phi 
\right ] \Psi(\phi) \; .
\end{equation}

The fundamental dynamical equation is the time-dependent functional
Schr\"odinger equation for the time-dependent state functional $\Psi(
\phi; t)$. This equation takes definite form, once a Hamiltonian
operator $H(\Pi, \Phi)$ is specified:

\begin{equation}
i \frac{\partial}{\partial t} \Psi(\phi; t ) =
H \left [  \frac{1}{i} \frac{\delta}{\delta \phi}, \phi 
\right ] \Psi(\phi; t) \; .
\label{eq:schr}
\end{equation}
In order to the initial value problem be completely defined, we must
also specify the initial wave functional.

In the case of a free scalar field it is rather simple make the
connection with the usual approaches to field theory in curved
space-time$^{\cite{ebo}}$. For instance, the vacuum is given by a
Gaussian wave functional which is not unique since there are many
orthogonal solutions to the equations of motion. Moreover, we can also
make field combination to define creation and annihilation operators to
discuss particle production in curved space-time. For further details see
refs.{\cite{ebo}},{\cite{sch}}.  

The time-dependent Schr\"odinger, Eq.\ (\ref{eq:schr}), cannot be
solved exactly, unless the system is described by a quadratic
Hamiltonian. To obtain a solution for a non-linear (interacting)
system we shall employ a variational approximation in which Dirac's
time-dependent variational principle$^{\cite{dir}}$ is implemented
approximately.

In classical mechanics the Newtonian equations are derived by
Hamiltonian's variational principle, which requires stationarizing the
classical action $I = \int dt~ L (\dot{q},q)$,

\begin{equation}
\frac{\delta I}{\delta q(t)} = 0 \; ,
\end{equation}
with the constraint that the variation $\delta q(t)$ vanishes at the
end points of the integral defining the action.

The quantum analogue of this principle for pure states is the Dirac's
variational principle, which leads to the time-dependent Schr\"odinger
equation.  First, we define the effective action $\Gamma$ as the time
integral of the diagonal matrix element of $i\partial /
\partial t -H$:

\begin{equation}
\Gamma = \int dt~ \left \langle \Psi \left | i \frac{\partial}{\partial t}
-H \right | \Psi \right \rangle \; ,
\label{dirac}
\end{equation}
and then we demand that $\Gamma$ be stationary against arbitrary
variations of $|\Psi \rangle$ and $\langle \Psi |$, subject to
appropriate boundary conditions, like it occurs in the Hamilton's
variational principle. Applications of this variational principle with
restricted variational {\em Ansatz}, in the Rayleigh-Ritz manner,
result in tractable self-consistent dynamical equations for the
parameters used in the {\em Ansatz}, which still retain some of the
non-linearity of the complete problem.


\section{The Model}
\label{model}


Many analyses of the time evolution of the scalar field associated to
inflation were carried out assuming that the space is homogeneous and
isotropic. In this work, we relax this hypothesis in order to assess
the importance of inhomogeneities. To accomplish this goal and keep the
problem tractable, we assumed that space is inhomogeneous only in one
direction, which we defined to be the $z$ direction.  This resulting 
space-time is the simplest inhomogeneous one. This scenario is
conveniently described by the Centrella-Wilson metric$^{\cite{suo}}$,

\begin{equation}
g_{\mu \nu}=\left(
\begin{array}{cccc}
 -{\sigma}^{2}+A^{2}{\beta^{x}}^{2}+A^{2}{\beta^{z}}^{2}    & A^{2}\beta^{x} & 
     0     & A^{2}\beta^{z} \\
 A^{2}\beta^{x}                                             &     A^{2}      & 
     0     &      0         \\
       0                                                    &        0       & 
A^{2}l^{2} &      0         \\
                          A^{2}\beta^{z}                    &        0       & 
     0     &     A^{2}
\end{array}
\right) \;\;,
\label{centrella}
\end{equation}
The parameters $A$, $\sigma$, $\beta^x$, $\beta^z$, and $l$ are
functions of the time $t$ and the coordinate $z$. The $\sigma$ term is 
a lapse function and it measures the orthogonal proper time interval 
$\sigma\delta t$ between neighboring slices a coordinate distance 
$\delta t$ apart. A shift vector, $\beta^{i}=\left(\beta^{x}, 0,
\beta^{z}\right)$, is used to simplify the three metric and keep it
diagonal. The three metric of the slices is $\gamma_{ij}$ and is take
to have the diagonal form

\begin{equation}
\gamma_{ij}=A^{2}\left(
\begin{array}{ccc}  
 1 & 0     & 0 \\
 0 & l^{2} & 0 \\
 0 & 0     & 1
\end{array}
\right) \;\;.
\end{equation}
The function $l$ is an anisotropic metric shear variable and the
conformal factor $A$ is a generalization of the usual scale factor $R$
in the homogeneous and isotropic Friedmann--Robertson-Walker (FRW)
models. Notice that the signature of the metric (\ref{centrella}) is
(-,+,+,+). The diagonal form of $\gamma_{ij}$ is preserved when the
coordinate trajectories are adjusted by using $\beta^{i}$, which results
in two nonzero shift vector components in the four metric
(\ref{centrella}).

The choice of this metric help us understand inhomogeneous, anisotropic
cosmologies in terms of a simple, homogeneous, isotropic FRW model. For
example, this metric generalizes the flat ($k=0$) FRW cosmologies, for
which $A(z,t)=R(t)$, $\sigma^{2}=1$, $\beta^{x} =\beta^{z}=0$ and
$l=1$. As we can easily see, the number parameters defining this metric
is substantially larger than the ones used to describe
Robertson--Walker space-times, even assuming such a minimal modification.

We assumed that the relevant degrees of freedom for the analyses of
inflation are described by a $\lambda \phi^4$ model, whose action is
given by

\begin{equation}                                                           
I=\int d^{n}x~{\cal L}=-\int d^{n}x \sqrt{-g}\left(\frac{1}{2}g^{\mu \nu}
\partial_{\mu}\Phi\partial_{\nu}\Phi+\frac{\mu^{2}}{2}\Phi^{2}
+\frac{\lambda}{4!}\Phi^{4}\right) \;\;,                     
\label{action}
\end{equation}                  
whereas $n=d+1$ with $d$ being the number of spatial dimensions. From
this action, we obtain that the canonical momentum conjugated to field 
$\Phi$ is

\begin{equation}
\Pi=\frac{\partial{\cal L}}{\partial\dot{\Phi}}=-\sqrt{-g}\left(g^{tt}
\dot{\Phi}+g^{ti}\partial_{i}\Phi\right) \;\;.
\end{equation}
Using the last two equations we can calculate the Hamiltonian density
of our model:

\begin{eqnarray}
{\cal H}(\Pi,\Phi)&=&\int_{\bf x}\left[ \frac{\sigma}{2\sqrt
                     {\epsilon}}\Pi^{2}+\sigma^{2}g^{it}\Pi\partial_{i}\Phi
                     +\frac{1}{2\sigma\sqrt{\epsilon}}\left(g^{ij}+\sigma^{2}
                     g^{it}g^{jt}\right)\partial_{i}\Phi\partial_{j}\Phi
                     \right. \nonumber\\
                  &+&\left. \frac{1}{\sigma\sqrt{\epsilon}}\left(\frac{
                     \mu^{2}}{2}\Phi^{2}+\frac{\lambda}{4!}
                     \Phi^{4}\right)\right] \;\;,
\label{hamiltonian}
\end{eqnarray}
where $\epsilon$ is $A^6l^2$.

We used the functional Schr\"odinger picture to describe the states and a
time-dependent variational approximation based on the Gaussian {\em
Ansatz}

\begin{eqnarray} 
\Psi( \phi, t) = && N(t) ~ \exp \left \{ i \int_{\bf x} \hat{\pi}({\bf x},t) 
\left [  \phi ({\bf x})  - \varphi ({\bf x},t) 
\right ] \right \}
\nonumber \\
&& \times \exp \left \{ -\int_{{\bf x},{\bf y}} 
[\phi({\bf x})-\varphi({\bf x},t)]
\left[ \frac{1}{4}\Omega^{-1}({\bf x},{\bf y},t)-i\Sigma
({\bf x}, {\bf y},t) \right]
\left[ \phi({\bf y})-\varphi({\bf y},t) \right ] 
\right \} \; ,
\label{gauss:pure}
\end{eqnarray}
where the variational parameters are $\varphi$, $\hat\pi$, $\Omega$,
and $\Sigma$ and we abbreviated the integral in $d$ spatial dimensions
as $\int_{\bf x} \equiv \int d^d{\bf x}$.  The physical meaning of the
parameters of this wave functional can be inferred from linear and
bilinear averages.  The linear averages are given by

\begin{eqnarray}
\langle \Phi({\bf x}) \rangle &=& \varphi({\bf x}, t) \;\;,
\\
\langle \Pi({\bf x}) \rangle &=& \hat{\pi}({\bf x}, t) \;\;,
\end{eqnarray}
while bilinear averages are
\begin{eqnarray}
&&\langle \Phi({\bf x}) \Phi({\bf y}) \rangle = \varphi({\bf x})
\varphi({\bf y}) + \Omega ({\bf x}, {\bf y}, t) \;\;, \\
&&\langle \Pi({\bf x}) \Pi({\bf y}) \rangle = \hat\pi({\bf x})
\hat\pi({\bf y}) + \frac{1}{4}\Omega^{-1} ({\bf x},{\bf y},t)+4
\left(\Sigma \Omega\Sigma \right)({\bf x},{\bf y},t) \;\;, \\
&&\langle \Phi({\bf x}) \Pi({\bf y}) \rangle = \frac{i}{2}
\delta({\bf x}-{\bf y}) + 2 \left (\Omega \Sigma \right )
({\bf x},{\bf y},t) \;\;,
\end{eqnarray}
where we have used the matrix notation $ \left({\cal O} {\cal
K}\right) ({\bf x}, {\bf y}) = \int_{{\bf z}} {\cal O}({\bf x}, {\bf
z}) {\cal K}({\bf z}, {\bf y})$.  Moreover, from the average value of
the operator $i (\partial/\partial t)$, appearing in the effective
action (\ref{dirac}), we find that the imaginary part of the
covariance function ($\Sigma$) plays the role of a canonical
momentum conjugate to the real part $\Omega$, and the pair
$(\pi,\varphi)$ has the same meaning,

\begin{equation}
\left \langle i \frac{\partial}{\partial t} \right \rangle =
\int_{\bf x} \hat\pi({\bf x}, t) \dot\varphi({\bf x}, t)
+ \int_{\bf x,y} \Sigma({\bf x}, {\bf y}, t) \dot\Omega
({\bf y}, {\bf x}, t) \;\;.
\end{equation}


\subsection*{Gaussian {\it Ansatz} in Curved Space}


In the spirit of Ref. \cite{hug}, we simplified the Gaussian {\it
Ansatz} specifying the momentum dependence of the kernels $\Omega$ and
$\Sigma$. This simplification is necessary not only to reduce the task
of proving the renormalizability of our model, but also to make
possible to apply it to numerical simulations. The kernels cam be
expressed as a Fourier transform (TF)

\begin{equation}
\Omega ({\bf x}, {\bf y}, t) = \int_{{\bf k}} e^{i{\bf k} \cdot ({\bf x}
- {\bf y})} \Omega({\bf k}, t) \; ,
\end{equation}
and we choose that

\begin{equation}
\Omega({\bf k},t)=\frac{1}{2\sqrt{\epsilon\left[A^{-2}\left(k_{x}^{2}
+l^{-2}k_{y}^{2}+k_{z}^{2}\right)+\alpha\right]}}\;\;, 
\label{ans:om}
\end{equation}

\begin{equation}
\Sigma({\bf k},t)=-\frac{\zeta}{8\left\{\epsilon\left[A^{-2}\left(k_{x}^{2}
+l^{-2}k_{y}^{2}+k_{z}^{2}\right)+\alpha\right]\right\}^{n}} \;\;.
\label{ans:sig}
\end{equation}
where the momentum-space integral $(\int~d^d{\bf k}/(2\pi)^d)$ is
denoted by $\int_{\bf k}$. Therefore, we trade the variational
parameters $\Omega$ and $\Sigma$ by $\alpha$ and $\zeta$, in order to
simplify our analysis. The parameter $n$ is constrained by the
renormalizability of the model, as we shall see below. It is
interesting to notice that this form for the kernels corresponds to
homogeneous fluctuations in the physical coordinates$^{\cite{hug}}$.
This is a reasonable hypothesis since the high frequency fluctuations
should be insensitive to large scale effects.

The Gaussian effective action is obtained through the substitution of
the {\em Ansatz} given by Eqs.\ (\ref{gauss:pure}), (\ref{ans:om}),
and (\ref{ans:sig}) into the Dirac's variational principle
(\ref{dirac}) with the Hamiltonian density (\ref{hamiltonian}),
\begin{eqnarray}                                                              
\Gamma &=&\int{dt} \int_{\bf x}\left\{\left[\hat{\pi}\dot{\varphi}
    -\frac{\sigma}{2\sqrt{\epsilon}}\hat{\pi}^{2}-\beta^{z}\hat{\pi}
    \partial_{z}\varphi-\sigma \sqrt{\epsilon}\left\{\frac{1}{2A^{2}}(\partial
    _{z}\varphi)^{2}-\frac{\mu^{2}}{2}\varphi^{2}
    -\frac{\lambda}{4!}\varphi^{4}\right\}\right]\right.
    \nonumber\\
&+& i\beta^{x}\left[\partial_{x}^{({\bf x})}\delta^{d}({\bf x}-{\bf y})
    \mid_{{\bf x}={\bf y}}-\int_{{\bf y}}\left\{\frac{1}{4}{\Omega}
    ^{-1}({\bf x},{\bf y}, t)+4i\Sigma({\bf x}, {\bf y},
    t)\right\}\partial _{x}^{({\bf x})}\Omega({\bf y},{\bf
    x},t)\right] 
    \nonumber\\
&+& i\beta^{z}\left[\partial_{z}^{({\bf x})}\delta^{d}({\bf x}-{\bf y})
    \mid_{{\bf x}={\bf y}}-\int_{{\bf y}}\left\{\frac{1}{4}{\Omega}^{-1}
    ({\bf x}, {\bf y}, t)+4i\Sigma({\bf x},{\bf y},t)\right\}
    \partial_{z}^{({\bf x})}\Omega({\bf y},{\bf x},t)\right] 
    \nonumber\\               
&+& \hbar \left[(\Sigma{\dot{\Omega}})({\bf x},{\bf x},t)-\frac{\sigma}
    {\sqrt{\epsilon}}\left\{\frac{1}{8}\Omega^{-1}({\bf x},{\bf x},t)+2(\Sigma
    {\Omega}\Sigma)({\bf x},{\bf x}, t)\right\} \right. 
    \nonumber\\
&+&\left.\frac{\sigma\sqrt{\epsilon}}{2A^{2}}\left(\left\{{\partial_{x}
   ^{({\bf x})}}^{2}+l^{-2}{\partial_{y}^{({\bf x})}}^{2}+{\partial_{z}
   ^{({\bf x})}}^{2}\right\}\Omega({\bf x},{\bf y}, t)\mid_{{\bf x}={\bf y}} 
   -\left(\mu^{2}+\frac{\lambda}{2}\varphi^{2}\right)\Omega({\bf
    x},{\bf x},t)\right)\right]
   \nonumber\\
&-&\left.\hbar^{2}\frac{\lambda\sigma \sqrt{\epsilon}}{8}\;\Omega
   ({\bf x}, {\bf x}, t)\Omega({\bf x},{\bf x}, t) \right\} \;\;.     
\label{act:curv}
\end{eqnarray}                                                                
Notice that we have set $\hbar=1$ throughout this paper except in
(\ref{act:curv}), in order to understand the structure of the Gaussian
approximation. Once more, the effective action have quantum
corrections up to $O\left({\hbar}^{2}\right)$, however, the
variational kernels are non-trivial functions of $\hbar$, which
re-introduces partially all orders in $\hbar$. In order to evaluate
the Gaussian effective action we considered $\varphi=\varphi(z,t)$,
and used the following expectation values,

\begin{eqnarray}
&&\langle \left(\partial_{j}\Phi\right)\Pi\rangle=-\hat{\pi}\partial_{j}
\varphi+2i\left(G\partial_{j}\Omega\right)({\bf x},{\bf x},t) \;\;, 
\label{expvalue1} \\
&&\langle \Pi\partial_{j}\Phi \rangle=-\hat{\pi}\partial_{j}\varphi
-i\partial_{j}^{({\bf x})}\delta^{d}({\bf x}-{\bf y})\mid_{{\bf x}={\bf y}}
+2i\left(G\partial_{j}\Omega\right)({\bf x},{\bf x},t) \;\;, 
\label{expvalue2} \\
&&\langle \partial_{i}\Phi \partial_{j}\Phi \rangle=\left(\partial_{i}\varphi
\right)\left(\partial_{j}\varphi\right)-\partial_{i}^{({\bf x})}\partial_{j}
^{({\bf y})}\Omega({\bf x},{\bf y},t)\mid_{{\bf x}={\bf y}} \;\;,
\label{expvalue3}
\end{eqnarray}
where we defined
\begin{equation}
G({\bf x},{\bf y},t)\equiv\frac{1}{4}\Omega^{-1}({\bf x},{\bf y},t)+4i\Sigma
({\bf x},{\bf y},t) \;\;.
\end{equation}

The stationary points of the effective action (\ref{act:curv}), with
respect to the variational parameters $\alpha$, $\zeta$, $\pi$, and
$\varphi$, leads to the equations of the motion of these parameters.
Notwithstanding, the Gaussian effective action exhibits infinities in
the limit $d \rightarrow 3$ due to the short distance behavior of
$\Omega$. Therefore, we must analyze its renormalization to obtain
finite equations of motion.


\section{Renormalization of Effective Action in the Centrella-Wilson Metric}


We regularized the divergences that appear in the effective action
(\ref{act:curv}) using dimensional regularization on the spatial
dimension $d$, since this is the natural procedure in the
Schr\"odinger picture$^{\cite{ebo}}$. In this section we show that this
effective action become finite by the renormalization
prescription$^{\cite{hug}}$

\begin{eqnarray} 
\frac{\mu^{2}}{\lambda}&=& \frac{\mu^{2}_{R}}{\lambda_{R}} \; ,
\label{mu:r} \\
\frac{1}{\lambda} &=& \frac{1}{\lambda_R} - 
\frac{2}{(4\pi)^{(d+1)/2} }~ \frac{1}{(d-1)(3-d)} \; .
\label{lambda:r} 
\end{eqnarray}

Initially we shall concentrate on the divergent terms of the effective
action (\ref{act:curv}), which are the ones involving the kernels
$\Omega$ and $\Sigma$. We divided the divergent terms into two sets
that are finite by themselves after the substitution of the
renormalization prescription. The first set contains the divergent
terms in the last three lines of the expression (\ref{act:curv}),

\begin{eqnarray}
D_{1}&=&-\frac{\sigma}{8\sqrt{\epsilon}}\Omega^{-1}({\bf x},{\bf x},t)
        +\frac{\sigma\sqrt{\epsilon}}{2A^{2}}\left[{\partial_{x}^
        {({\bf x})}}^{2}+l^{-2}{\partial_{y}^{({\bf x})}}
        ^{2}+{\partial_{z}^{({\bf x})}}^{2}\right]\Omega
        ({\bf x},{\bf y},t)\mid_{{\bf x}={\bf y}}
        \nonumber\\
     &-&\frac{\sigma\sqrt{\epsilon}}{2}\left(\mu^{2}+
        \frac{\lambda}{2}\varphi^{2}\right)\Omega
        ({\bf x},{\bf x},t)-\frac{\lambda\sigma\sqrt{\epsilon}}{8}
        \Omega({\bf x},{\bf x},t)\Omega({\bf x},{\bf x},t) \;\;.
\label{div:one}
\end{eqnarray}
In order to obtain the divergences in $D_1$, we used the Fourier
transform of the kernels, leading to
\begin{eqnarray}
D_{1}&=&-\frac{\sigma}{4}\left\{\int_{{\bf k}}\sqrt{A^{-2}\left(k_{x}^{2}
        +l^{-2}k_{y}^{2}+k_{z}^{2}+\right)+\alpha}+\frac{1}{A^{2}}
        \int_{{\bf k}}\frac{\left(k_{x}^{2}+l^{-2}k_{y}^{2}
        +k_{z}^{2}\right)}{\sqrt{A^{-2}\left(k_{x}^{2}+l^{-2}k_{y}^{2}
        +k_{z}^{2}\right)+\alpha}} \right. 
        \nonumber\\
     &+&\frac{\lambda}{8\sqrt{\epsilon}}\int_{{\bf k}}\frac{1}
        {\sqrt{A^{-2}\left(k_{x}^{2}+l^{-2}k_{y}^{2}+k_{z}^{2}\right)
        +\alpha}}\int_{{\bf k'}}\frac{1}{\sqrt{A^{-2}\left({k'_{x}}^{2}
        +l^{-2}{k'_{y}}^{2}+{k'_{z}}^{2}\right)+\alpha}} 
        \nonumber\\
     &+&\left.\left(\mu^{2}+\frac{\lambda}{2}\varphi^{2}\right)
        \int_{{\bf k}}\frac{1}{\sqrt{A^{-2}\left(k_{x}^{2}+l^{-2}k_{y}^{2}
        +k_{z}^{2}\right)+\alpha}}\right\} \;\;.
\label{div:int}
\end{eqnarray}
The above expression is written in terms of {\it co-moving}
coordinates. However, it is convenient to express it in the physical space
variables,
\begin{equation}
p_{x}=k_{x}/A\;\;,  \;\;\;\;
p_{y}=k_{y}/Al\;\;, \;\;\;\; 
p_{z}=k_{z}/A\;\;,
\label{trans}
\end{equation}
that leads to
\begin{eqnarray}
D_{1}&=&-\frac{\sigma\sqrt{\epsilon}}{2}\left\{\int_{{\bf p}}\sqrt{p^{2}
        +\alpha}-\frac{1}{2}\int_{{\bf p}}\frac{\alpha}{\sqrt{p^{2}+\alpha}} 
        +\frac{1}{2}\left(\mu^{2}+\frac{\lambda}{2}\varphi^{2}\right)
        \int_{{\bf p}}\frac{1}{\sqrt{p^{2}+\alpha}} \right. 
        \nonumber\\
     &+&\left. \frac{\lambda}{16}\int_{{\bf p},{\bf p'}}\frac{1}{\sqrt{p^{2}
        +\alpha}}\frac{1}{\sqrt{p'^{2}+\alpha}}\right\} \;\;.
\label{div:trans}
\end{eqnarray}
Using dimensional regularization$^{\cite{col}}$ to evaluate the above 
expression we obtain
\begin{eqnarray}
D_{1}&=&\frac{\sigma\sqrt{\epsilon}}{(4\pi)^{(d+1)/2}(1-d)}\left(
        \frac{\alpha}{\Lambda^2}\right)^{(d-3)/2} 
        \Gamma\left(\frac{3-d}{2}\right)\left\{\frac{(d-1)}
        {(d+1)}\alpha^{2}-\left(\mu^{2}+\frac{\lambda}{2} 
        {\varphi^{2}}\right)\alpha \right.
        \nonumber\\
     &-&\left. \frac{\lambda\alpha^{2}}{2(4\pi)^{(d+1)/2}(1-d)}
        \left(\frac{\alpha}{\Lambda^2}\right)^{(d-3)/2}\Gamma 
        \left(\frac{3-d}{2}\right)\right\} \;\; ,
\label{div:reg}
\end{eqnarray}    
where $\Lambda$ is an arbitrary mass scale which we choose to be given by

\begin{equation} 
\Lambda^{2}=\frac{\mu_{R}^{2}e^{-(\gamma-1/2)}}{4\pi} \; .
\label{Lam} 
\end{equation}
Substituting the renormalization prescription (\ref{mu:r}) --
(\ref{lambda:r}) into (\ref{div:reg}), we obtain $D_{1}$ in terms of
renormalized quantities $\mu_{R}^{2}$ and $\lambda_{R}$, which is
finite in the limit $d\rightarrow 3$:

\begin{eqnarray}
D_{1}^{R}=-\sigma\sqrt{\epsilon}\left\{\frac{\alpha}{2}\varphi^{2}
          +\frac{\alpha^{2}}{64\pi^{2}}\left[\ln\left(\frac{\alpha}
          {\mu_{R}^{2}}\right)-\frac{1}{2}\right]
          -\frac{\left(\alpha-\mu_{R}^{2}\right)^{2}}{2\lambda_{R}}
          +\frac{\mu_{R}^{4}}{2\lambda_{R}}\right\} \;\;.
\label{div1:res}
\end{eqnarray}
Notice that the above expression has a non-trivial dependence on
$\lambda_R$ due to the renormalization prescription that we used.

The second set of potentially divergent terms appearing in
(\ref{act:curv}) is
\begin{eqnarray}
D_{2}&=& i\beta^{x}\left[\partial_{x}^{({\bf x})}\delta^{d}({\bf x}-{\bf y})
    \mid_{{\bf x}={\bf y}}-\int_{{\bf y}}\left\{\frac{1}{4}{\Omega}^{-1}
    ({\bf x},{\bf y},t)+4i\Sigma({\bf x}, {\bf y},
    t)\right\}\partial_{x}^{({\bf y})}\Omega({\bf y},{\bf x},t)\right] 
    \nonumber\\
&+& i\beta^{z}\left[\partial_{z}^{({\bf x})}\delta^{d}({\bf x}-{\bf y})
    \mid_{{\bf x}={\bf y}}-\int_{{\bf y}}\left\{\frac{1}{4}{\Omega}^{-1}
    ({\bf x},
    {\bf y}, t)+4i\Sigma({\bf x}, {\bf y}, t)\right\}\partial_{z}^{({\bf y})}
    \Omega({\bf y}, {\bf x}, t)\right] \nonumber\\
&-& \frac{2\sigma}{\sqrt{\epsilon}}(\Sigma\Omega\Sigma)({\bf x},{\bf x},t)
    +\left(\Sigma\dot{\Omega}\right)({\bf x},{\bf x},t) \;\;.   
\label{div:two}
\end{eqnarray}
The Fourier transform of (\ref{div:two}), after the substitution of
the {\it Ans\"atze} (\ref{ans:om}) and (\ref{ans:sig}) can be written
in terms of the physical momenta as
\begin{eqnarray}
D_{2}&=&\sqrt{\epsilon}\int_{{\bf p}}\left\{A\left[i-\frac{1}{2}-\frac{i\zeta}
        {16\;\epsilon^{n}\left(p^{2}+\alpha\right)^{n+1/2}}\right]\left(\beta
        ^{x}p_{x}+\beta^{z}p_{z}\right)-\frac{\sigma\zeta^2}{64\epsilon
        ^{2n+3/2}\left(p^{2}+\alpha\right)^{2n+1/2}}\right.
        \nonumber\\
     &+&\left.\frac{\zeta}{32\epsilon^{n}}\left[\left(\dot{\alpha}
        +\alpha\left\{\frac{2\dot{A}}{A}+\frac{2\dot{l}}{3l}\right\}\right)
        \frac{1}{\left(p^{2}+\alpha\right)^{n+3/2}}+\left(\frac{\dot
        {\epsilon}}{\epsilon}-\frac{2\dot{A}}{A}-\frac{2\dot{l}}{3l}\right)
        \frac{1}{\left(p^2+\alpha\right)^{n+1/2}}\right]\right\} \;\;,
\label{div2:int}
\end{eqnarray}
where we used that
\begin{eqnarray}
\dot{\Omega}({\bf x},{\bf x},t)&=&\int_{{\bf k}}\dot{\Omega}({\bf k},t)
          \nonumber\\
       &=&-\frac{1}{4}\left\{\left(\frac{\dot{\epsilon}}{\epsilon}
          -\frac{2\dot{A}}{A}-\frac{2\dot{l}}{3l}\right)\int_{{\bf p}}
          \frac{1}{\sqrt{p^{2}+\alpha}}+\left[\dot{\alpha}+\alpha\left(
          \frac{2\dot{A}}{A}+\frac{2\dot{l}}{3l}\right)\right]\int_{{\bf
					p}}\frac{1}{\left(p^{2}+\alpha\right)^{3/2}}\right\}\;\;.
\end{eqnarray}

In dimensional regularization integrals of the form $\int_{{\bf p}}
p_{i}$ also vanish$^{\cite{col}}$, while the well-known result 
\begin{equation}
\int_{{\bf p}}\frac{p_{\mu}}{\left(p^{2}+2p\cdot P-M^{2}\right)^{\alpha}}=
\frac{-i\pi^{n/2}P_{\mu}}{(2\pi)^{d}\Gamma(\alpha)\left(-P^{2}-M^{2}\right)
^{\alpha-n/2}}\Gamma\left(\alpha-\frac{n}{2}\right) \;\;,
\end{equation}
shows that the above integrals $\int_{{\bf
p}}~p_{i}/\left(p^{2}+\alpha\right)^{n+1/2}$ also vanish since they
correspond to the case $P=0$. Now the only potentially divergent terms
in $D_2$ depend upon $n$. Since $D_2$ does not contain $\mu$ or 
$\lambda$, that could absorb a divergence, we choose $n > 2$
in $\Sigma$, in order to avoid new divergences in $D_2$.
Consequently, (\ref{div2:int}) is finite in the limit $d \rightarrow
3$ and given by
\begin{equation}
D_{2}^{{\rm finite}}=-\frac{\sigma\zeta^{2}}{64\epsilon^{2n+1}}
                     I_{2n+1/2}+\frac{\zeta}{32\epsilon^{n}}
                     \left\{\left[\dot{\alpha}+\alpha\left(\frac{2\dot{A}}{A}
                     +\frac{2\dot{l}}{3l}\right)\right]I_{n+3/2} 
                     +\left(\frac{\dot{\epsilon}}{\epsilon}
                     -\frac{2\dot{A}}{A}-\frac{2\dot{l}}{3l}\right)I_{n+1/2}
                     \right\} \;\;,
\label{div:2,2}
\end{equation}
where we defined
\begin{equation}
I_{n}\equiv \int_{\bf p}\frac{1}{\left(p^{2}+\alpha\right)^{n}} \;\;. 
\end{equation}

Finally, substituting (\ref{div1:res}) and (\ref{div:2,2}) into
(\ref{act:curv}) we obtain the renormalized effective action
\begin{eqnarray}
\Gamma_{R}&=&\int dt\int_{{\bf x}}\left\{\hat{\pi}\dot{\varphi}-\frac{\sigma
             \hat{\pi}^{2}}{2\sqrt{\epsilon}}+\beta^{z}\hat{\pi}\partial_{z}
             \varphi-\frac{\sigma\sqrt{\epsilon}}{2A^{2}}\left(\partial_{z}
             \varphi\right)^{2}+D_{2}^{{\rm finite}}+D_{1}^{R}\right\} 
             \nonumber\\
          &=&\int dt\int_{{\bf x}}\left\{\hat{\pi}\dot{\varphi}
             -\frac{\sigma\hat{\pi}^{2}}{2\sqrt{\epsilon}}
             +\beta^{z}\hat{\pi}\partial_{z}\varphi-\frac{\sigma\sqrt
             {\epsilon}}{2A^{2}}\left(\partial_{z}\varphi\right)^{2}
             -\frac{\sigma{\zeta}^{2}}{64{\epsilon}^{2n+1}}I_{2n+1/2}\right.
             \nonumber\\
          &+&\frac{\zeta}{32{\epsilon}^{n}}
             \left[\left(\dot{\alpha}+\alpha\left\{\frac{2\dot{A}}{A}+\frac{2
             \dot{l}}{3l}\right\}\right)I_{n+3/2}+\left(\frac{
             \dot{\epsilon}}{\epsilon}-\frac{2\dot{A}}{A}-\frac{2\dot{l}}{3l}
             \right)I_{n+1/2}\right]
             \nonumber\\
          &-&\left.\sigma\sqrt{\epsilon}\left[\frac{1}{2}{\varphi}^{2}\alpha
             +\frac{1}{64\pi^{2}}
             {\alpha}^{2}\left\{\ln\left(\frac{\alpha}{\mu_{R}^{2}}\right)
             -\frac{1}{2}\right\}-\frac{\left(\alpha-\mu_{R}^{2}\right)^{2}}
             {2\lambda_{R}}+\frac{\mu_{R}^{4}}{2\lambda_{R}}\right]
             \right\} \;\;.
\label{act:ren}
\end{eqnarray}
At this point we can obtain the finite (renormalized) equations of
motion for the variational parameters $\hat{\pi},\varphi,\zeta$ and
$\alpha$, by varying (\ref{act:ren}) with respect to them.
\begin{eqnarray}
&&\frac{\delta \Gamma_{R}}{\delta \varphi}=0 \Rightarrow \dot{\hat{\pi}}=
-\sigma\sqrt{\epsilon}~\alpha\varphi \;\;, \\
&&\frac{\delta \Gamma_{R}}{\delta \hat{\pi}}=0 \Rightarrow
\dot{\varphi}=\beta^{z}\partial_{z}\varphi-\frac{\sigma\hat{\pi}}{2
\sqrt{\epsilon}}\;\;, \\
&&\frac{\delta \Gamma_{R}}{\delta \zeta}=0 \Rightarrow
\dot{\alpha}=I^{-1}_{n+3/2}\left\{\frac{\sigma\zeta}{{\epsilon}^{n+1}}
I_{2n+1/2}-\alpha\left(\frac{2\dot{A}}{A}+\frac{2\dot{l}}{3l}\right)I_{n+3/2}
-\left(\frac{\dot{\epsilon}}{\epsilon}-\frac{2\dot{A}}{A}-\frac{2\dot{l}}
{3l}\right)I_{n+1/2}\right\} \;\;,
\end{eqnarray}
\begin{eqnarray}
 \frac{\delta \Gamma_{R}}{\delta \alpha}=0 \Rightarrow \dot{\zeta} 
  &=& I^{-1}_{n+3/2}\left\{\zeta\left(n+\frac{1}{2}\right)
      \left(\frac{\dot{\epsilon}}{\epsilon}-\frac{2\dot{A}}{A}
      -\frac{2\dot{l}}{3l}\right)I_{n+3/2}\right.
      \nonumber\\
  &+& \zeta\left(n+\frac{3}{2}\right)\left[\dot{\alpha}+\alpha\left(
      \frac{2\dot{A}}{A}+\frac{2\dot{l}}{3l}\right)\right]I_{n+5/2}
      \nonumber\\
  &-& \left. 32\sigma{\epsilon}^{n+1/2}\left[\frac{1}{2}\varphi^{2}
      +\frac{1}{32\pi^{2}}\alpha\ln\left(\frac{
      \alpha}{\mu_{R}^{2}}\right)-\frac{(\alpha-\mu_{R}^{2})}{\lambda_{R}}
      \right]\right\} \;\;.
\end{eqnarray}


\section{Renormalization of Energy-Momentum Tensor in Inhomogeneous Space}


The variation of the action (\ref{action}) with respect to the metric
tensor $g^{\mu \nu}$, results in the energy-momentum tensor
\begin{equation}
T_{\mu \nu}=\partial_{\mu}\Phi \partial_{\nu}\Phi-g_{\mu \nu}\left(\frac{1}
              {2}g^{\alpha \beta}\partial_{\alpha}\Phi \partial_{\beta}\Phi
              +\frac{\mu^{2}}{2}\Phi^{2}+\frac{\lambda}{4!}\Phi^{4}\right) 
              \;\;.
\label{tensor:energ}
\end{equation}
In this section we shall analyze the renormalization of the expectation
value of $T_{\mu\nu}$ in our Gaussian {\em Ansatz}. 


\subsection{Renormalization of Diagonal Components of Energy-Momentum Tensor}


The expectation value of $T_{\mu \nu}$, for $\mu=\nu$, in the
Gaussian state (\ref{gauss:pure}) is given by
\begin{eqnarray}
\langle T_{tt} \rangle&=&\frac{{\hat{\pi}}^{2}}{\epsilon}+{\beta^{z}}^{2}\left
         (\partial_{z}\varphi\right)^{2}+\frac{2\sigma\beta^{z}}
         {\sqrt{\epsilon}}\hat{\pi}\partial_{z}\varphi 
         \nonumber\\
      &-&\left(-\sigma^{2}+A^{2}{\beta^{x}}^{2}+A^{2}{\beta^{z}}^{2}\right)
         \left[-\frac{{\hat{\pi}}^{2}}{2\epsilon}+\frac{1}{2}\left(\partial
         _{z}\varphi\right)^{2}+\frac{\mu^{2}}{2}\varphi^{2}+\frac{\lambda}{4!}
         \varphi^{4}\right] \nonumber\\
      &-&\frac{\sigma}{\sqrt{\epsilon}}\beta^{x}\left[-i\partial_{x}
         ^{({\bf x})}\delta^{d}({\bf x}-{\bf y})\mid_{{\bf x}={\bf y}}
         +4i\left(G\partial_{x}\Omega\right)({\bf x},{\bf x},t)\right]
         \nonumber\\
      &-&\frac{\sigma}{\sqrt{\epsilon}}\beta^{z}\left[-i\partial_{z}
         ^{({\bf x})}\delta^{d}({\bf x}-{\bf y})\mid_{{\bf x}={\bf y}}
         +4i\left(G\partial_{z}\Omega\right)({\bf x},{\bf x},t)\right]
         \nonumber\\
      &+&\left(-\sigma^{2}+A^{2}{\beta^{x}}^{2}+A^{2}{\beta^{z}}^{2}\right)
         \left\{\frac{1}{2A^{2}}\left[{\partial_{x}^{({\bf x})}}^{2}+l^{-2}
         {\partial_{y}^{({\bf x})}}^{2}+{\partial_{z}^{({\bf x})}}^{2}\right]
         \Omega({\bf x},{\bf y},t)\mid_{{\bf x}={\bf y}}\right. \nonumber\\
      &-&\left. \frac{1}{8\epsilon}\Omega^{-1}({\bf x},{\bf x},t)-\frac{1}{2}
         \left(\mu^{2}+\frac{\lambda}{2}{\varphi}^{2}\right)\Omega({\bf x},
         {\bf x},t)-\frac{\lambda}{8}\Omega({\bf x},{\bf x},t)\Omega({\bf x},
         {\bf x},t)\right\} \nonumber\\
      &-&\left[{\beta^{x}}^{2}{\partial_{x}^{({\bf x})}}^{2}+{\beta^{z}}^{2}
         {\partial_{z}^{({\bf x})}}^{2}+2\beta^{x}\beta^{z}\partial_{x}^
         {({\bf x})}\partial_{z}^{({\bf x})}\right]\Omega({\bf x},{\bf y},t)
         \mid_{{\bf x}={\bf y}}+\frac{\sigma^{2}}{4\epsilon}\Omega^{-1}
         ({\bf x},{\bf x},t) \nonumber\\
      &+&\frac{2}{\epsilon}\left(\sigma^{2}+A^{2}{\beta^{x}}^{2}+A^{2}{\beta
         ^{z}}^{2}\right)(\Sigma\Omega\Sigma)({\bf x},{\bf x},t) \;\;,
\label{Ttt:esp}
\end{eqnarray}

\begin{eqnarray}
\langle T_{xx} \rangle &=&-A^{2}\left\{-\frac{{\hat{\pi}}^{2}}{2\epsilon}
         +\frac{{\beta^{z}}^{2}}{2A^{2}}\left(\partial_{z}\varphi\right)^{2}
         +\frac{\mu^{2}}{2}\varphi^{2}+\frac{\lambda}{4!}\varphi^{4}\right.
         \nonumber\\
      &-&\frac{1}{2A^{2}}\left[{\partial_{x}^{({\bf x})}}^{2}+l^{-2}{\partial
         _{y}^{({\bf x})}}^{2}+{\partial_{z}^{({\bf x})}}^{2}\right]
         \Omega({\bf x},{\bf y},t)\mid_{{\bf x}={\bf y}}+\frac{1}{8\epsilon}
         \Omega^{-1}({\bf x},{\bf x},t) \nonumber\\
      &+&\frac{1}{2}\left(\mu^{2}+\frac{\lambda}{2}{\varphi}^{2}\right)\Omega
         ({\bf x},{\bf x},t)+\frac{\lambda}{8}\Omega({\bf x},{\bf x},t)
         \Omega({\bf x},{\bf x},t) \nonumber\\
      &-&\left. \frac{2}{\epsilon}(\Sigma\Omega\Sigma)({\bf x},{\bf x},t)
         \right\}-{\partial_{x}^{({\bf x})}}^{2}\Omega({\bf x},{\bf y},t)\mid
         _{{\bf x}={\bf y}} \;\;,     
\label{Txx:esp}
\end{eqnarray}

\begin{eqnarray}
\langle T_{yy} \rangle &=&-(Al)^{2}\left\{-\frac{{\hat{\pi}}^{2}}{2\epsilon}
         +\frac{{\beta^{z}}^{2}}{2A^{2}}\left(\partial_{z}\varphi\right)^{2}
         +\frac{\mu^{2}}{2}\varphi^{2}+\frac{\lambda}{4!}\varphi^{4}\right. 
         \nonumber\\
      &-&\frac{1}{2A^{2}}\left[{\partial_{x}^{({\bf x})}}^{2}+l^{-2}{\partial
         _{y}^{({\bf x})}}^{2}+{\partial_{z}^{({\bf x})}}^{2}\right]
         \Omega({\bf x},{\bf y},t)\mid_{{\bf x}={\bf y}}+\frac{1}{8\epsilon}
         \Omega^{-1}({\bf x},{\bf x},t) \nonumber\\
      &+&\frac{1}{2}\left(\mu^{2}+\frac{\lambda}{2}{\varphi}^{2}\right)\Omega
         ({\bf x},{\bf x},t)+\frac{\lambda}{8}\Omega({\bf x},{\bf x},t)
         \Omega({\bf x},{\bf x},t) \nonumber\\
      &-&\left. \frac{2}{\epsilon}(\Sigma\Omega\Sigma)({\bf x},{\bf x},t)
         \right\}-l^{-2}{\partial_{y}^{({\bf x})}}^{2}\Omega({\bf x},{\bf y},t)
         \mid_{{\bf x}={\bf y}} \;\;, 
\label{Tyy:esp}
\end{eqnarray}

\begin{eqnarray}
\langle T_{zz} \rangle
&=&\left(\partial_{z}\varphi\right)^{2}-A^{2}\left\{-\frac{{\hat{\pi}}^{2}}
         {2\epsilon}+\frac{{\beta^{z}}^{2}}{2A^{2}}\left(\partial_{z}
         \varphi\right)^{2}+\frac{\mu^{2}}{2}\varphi^{2}+\frac{\lambda}{4!}
         \varphi^{4}\right.
         \nonumber\\
      &-&\frac{1}{2A^{2}}\left[{\partial_{x}^{({\bf x})}}^{2}+l^{-2}{\partial
         _{y}^{({\bf x})}}^{2}+{\partial_{z}^{({\bf x})}}^{2}\right]
         \Omega({\bf x},{\bf y},t)\mid_{{\bf x}={\bf y}}+\frac{1}{8\epsilon}
         \Omega^{-1}({\bf x},{\bf x},t) \nonumber\\
      &+&\frac{1}{2}\left(\mu^{2}+\frac{\lambda}{2}{\varphi}^{2}\right)\Omega
         ({\bf x},{\bf x},t)+\frac{\lambda}{8}\Omega({\bf x},{\bf x},t)
         \Omega({\bf x},{\bf x},t) \nonumber\\
      &-&\left. \frac{2}{\epsilon}(\Sigma\Omega\Sigma)({\bf x},{\bf x},t)
         \right\}-{\partial_{z}^{({\bf x})}}^{2}\Omega({\bf x},{\bf y},t)
         \mid_{{\bf x}={\bf y}} \;\;. 
\label{Tzz:esp}
\end{eqnarray}
The divergent terms that are common to the four diagonal components
are given by
\begin{eqnarray}
D_{\mu \mu}&\equiv &g_{\mu \mu}\left\{\frac{1}{2A^{2}}\left[{\partial_
              {x}^{({\bf x})}}^{2}+l^{-2}{\partial_{y}^{({\bf x})}}^{2}
              +{\partial_{z}^{({\bf x})}}^{2}\right]\Omega({\bf x}, 
              {\bf y},t)\mid_{{\bf x}={\bf y}}-\frac{1}{8\epsilon}
              \Omega^{-1}({\bf x},{\bf x},t)\right. \nonumber\\
           &-&\left.\frac{1}{2}\left(\mu^{2}+\frac{\lambda}{2}\varphi^{2}
              \right)\Omega({\bf x}, {\bf x},t)-\frac{\lambda}{8}\Omega
              ({\bf x},{\bf x},t)\Omega({\bf x},{\bf x},t)\right\}\;\;,
\label{Ttt:div}
\end{eqnarray}
where there is no sum in the $\mu$ index. Substituting the {\it
Ansatz} (\ref{ans:om}) and using the physical momenta (\ref{trans}) results in
\begin{eqnarray}
D_{\mu \mu}&=&g_{\mu \mu}\left\{\frac{\alpha}{4}-\frac{1}{4}\left(\mu^{2}
              +\frac{\lambda}{2}\varphi^{2}\right)-\frac{\lambda}{32}
              \int_{{\bf p'}}\frac{1}{\sqrt{p'^{2}+\alpha}}\right\}
              \int_{\bf p}\frac{1}{\sqrt{p^{2}+\alpha}} \;\;.
\end{eqnarray}
At this point we evaluate this expression using dimensional
regularization in order to extract the divergences,
\begin{eqnarray}
D_{\mu\mu}&=&\frac{g_{\mu\mu}}{(4\pi)^{(d+1)/2}(1-d)}\left(\frac{\alpha}
             {\Lambda^2}\right)^{(d-3)/2}\Gamma\left(\frac{3-d}{2}\right)
             \left\{\alpha^{2}-\left(\mu^{2}+\frac{\lambda}{2}{\varphi^{2}}
             \right)\alpha\right.
             \nonumber\\
          &-&\left. \frac{\lambda\alpha^{2}}{2(4\pi)^{(d+1)/2}(1-d)}
             \left(\frac{\alpha}{\Lambda^2}\right)^{(d-3)/2}\Gamma\left(
             \frac{3-d}{2}\right)\right\} \;\;.
\end{eqnarray}
Notice that the $\langle\Phi^{4} \rangle$ term gives rise to a double
pole since its expectation value is proportional to the square of
$\Omega({\bf x}, {\bf x},t)$. However, this double pole comes
multiplied by $\lambda$, which transform the double pole into a single
one due to (\ref{lambda:r}).              

Substituting the renormalization prescription (\ref{mu:r}) --
(\ref{lambda:r}) for $\mu^2$ and $\lambda$, we can see that the the
only terms that survive in the limit $d\rightarrow 3$ are
\begin{equation}
D_{\mu \mu}=-\frac{g_{\mu \mu}}{32\pi^{2}}\left[\frac{\alpha^{2}}{(3-d)} 
+32\pi^{2}\alpha\left(\frac{\mu_{R}^{2}}{\lambda_{R}}+\frac{1}{2}\varphi^{2}
\right)-16\pi^{2}\frac{\alpha^{2}}{\lambda_{R}}\right] \;\;,
\label{Dmumu:polo}
\end{equation}
which still contains a pole. Nevertheless, there are other divergences
in (\ref{Ttt:esp}) -- (\ref{Tzz:esp}) that should be added to this one. 
In the case of $\langle T_{tt} \rangle$, the remaining divergent terms are

\begin{equation} 
D_{3}\equiv-\left[{\beta^{x}}^{2}{\partial_{x}^{({\bf
x})}}^{2}+{\beta^{z}}^{2}{\partial_ {z}^{({\bf
x})}}^{2}\right]\Omega({\bf x},{\bf y},t)\mid_{{\bf x}={\bf y}}
+\frac{\sigma^{2}}{4\epsilon}\Omega^{-1}({\bf x},{\bf x},t) \;\;.
\label{T:div2}
\end{equation}
We already knows to calculate the terms in (\ref{T:div2}), adopting
the same procedure used in the previous section,  

\begin{equation}
D_{3}=\frac{\alpha^{2}}{32\pi^{2}}\left(-\sigma^{2}+A^{2}{\beta^{x}}^{2}+A^{2}
{\beta^{z}}^{2}\right)\left\{\frac{1}{(3-d)}-\frac{1}{2}\left[\ln
\left(\frac{\alpha}{\mu_{R}^{2}}\right)-\frac{1}{2}\right]\right\} \;\;.
\label{D3:polo}
\end{equation}
Summing (\ref{Dmumu:polo}), (\ref{D3:polo}), we can to see that the 
coefficient of the pole in the limit $d\rightarrow 3$ is canceled, 
when $\mu=\nu=t$:

\begin{eqnarray}
D_{tt}+D_{3}&=&\left(-\sigma^{2}+A^{2}{\beta^{x}}^{2}+A^{2}{\beta^{z}}
               ^{2}\right)\left\{-\frac{\alpha}{2}\varphi^{2} 
               -\frac{\alpha^{2}}{64\pi^{2}}\left[\ln
               \left(\frac{\alpha}{\mu_{R}^{2}}\right)-\frac{1}{2}\right]
               \right. \nonumber\\
            &+&\left.\frac{\left(\alpha-\mu_{R}^{2}\right)^{2}}
               {2\lambda_{R}}-\frac{\mu_{R}^{4}}{2\lambda_{R}}\right\} \;\;.
\end{eqnarray}
Here it is easy to verify that the remaining terms in (\ref{Ttt:esp})
are finite, which leads to the following for $\langle T_{tt} \rangle$ 
in terms of renormalized coupling constants $\mu_{R}^{2}$ and $\lambda_{R}$:

\begin{eqnarray}
\langle T_{tt}\rangle_{R}&=&\frac{{\hat{\pi}}^{2}}{\epsilon}
                         +{\beta^{z}} ^{2}\left(\partial_{z}
                         \varphi\right)^{2}+\frac{2\sigma\beta^{z}}{\sqrt{
                         \epsilon}}\hat{\pi}\partial_{z}\varphi \nonumber\\
                      &-&\left(-\sigma^{2}+A^{2}{\beta^{x}}^{2}+A^{2}
                         {\beta^{z}}^{2}\right)\left\{
                         -\frac{{\hat{\pi}}^{2}}{2\epsilon}
                         +\frac{1}{2}\left(\partial_{z}\varphi
                         \right)^{2}+\frac{\alpha}{2}\varphi^{2}\right. 
                         \nonumber\\
                      &+&\left.\frac{\alpha^{2}}{64\pi^{2}}\left[\ln\left(
                         \frac{\alpha}{\mu_{R}^{2}}\right)+\frac{1}{2}\right]
                         -\frac{\left(\alpha-\mu_{R}^{2}\right)^{2}}{2\lambda
                         _{R}}+\frac{\mu_{R}^{4}}{2\lambda_{R}}\right\}   
                         \nonumber\\
                      &+&\left(\sigma^{2}+A^{2}{\beta^{x}}^{2}+A^{2}{\beta^{z}}
                         ^{2}\right)\frac{\zeta^{2}}{64{\epsilon}^{2n+1}}
                         I_{2n+1/2} \;\;.     
\end{eqnarray}
The divergences in $\langle T_{ij} \rangle$, when $i=j$, can be written
as 

\begin{equation}
D_{ii}-\frac{g_{ii}}{A^{2}}{\partial_{i}^{({\bf x})}}^{2}\Omega(
{\bf x},{\bf y},t)\mid_{{\bf x}={\bf y}} \;\;,
\label{term:Dii}
\end{equation}
where there is no sum in the $i$ index. The Fourier transform of second
term in (\ref{term:Dii}) is given by

\begin{equation}
\frac{g_{ii}}{A^{2}}{\partial_{i}^{({\bf x})}}^{2}\Omega({\bf x},{\bf y},t)
\mid_{{\bf x}={\bf y}}=-\frac{g_{ii}}{2d}\int_{{\bf p}}\frac{p^{2}}
{\sqrt{p^{2}+\alpha}} \;\;,
\end{equation}
whose value in the limit $d\rightarrow 3$ is

\begin{equation}
-\frac{g_{ii}}{2d}\int_{{\bf p}}\frac{p^{2}}{\sqrt{p^{2}+\alpha}}=
-\frac{g_{ii}\alpha^{2}}{32\pi^{2}}\left\{\frac{1}{(3-d)}-\frac{1}{2}
\left[\ln\left(\frac{\alpha}{\mu_{R}^{2}}\right)-\frac{1}{2}\right]
\right\} \;\;.
\label{rest:finit}
\end{equation}
In this way, the coefficient of the pole in (\ref{term:Dii}) vanishes,
resulting in

\begin{eqnarray}
D_{ii}-\frac{g_{ii}}{A^{2}}{\partial_{i}^{({\bf x})}}^{2}\Omega({\bf x},
{\bf y},t)\mid_{{\bf x}={\bf y}}
         &=&g_{ii}\left\{-\frac{\alpha}{2}\varphi^{2}-\frac{\alpha^{2}}
            {64\pi^{2}}\left[\ln\left(\frac{\alpha}{\mu_{R}^{2}}\right)
            -\frac{1}{2}\right] \right. \nonumber\\
         &+&\left. \frac{\left(\alpha-\mu_{R}^{2}\right)^{2}}{2\lambda_{R}}
            -\frac{\mu_{R}^{4}}{2\lambda_{R}}\right\} \;\;. 
\end{eqnarray}
The spatial diagonal components of $\langle T_{\mu \nu} \rangle$ in
terms of renormalized constants coupling $\mu_{R}^{2}$ and
$\lambda_{R}$, are given by

\begin{eqnarray}
\langle T_{xx} \rangle_{R}&=&-A^{2}\left\{-\frac{{\hat{\pi}}^{2}}{2\epsilon}
                         +\frac{{\beta^{z}}^{2}}{2A^{2}}\left(\partial_{z}
                         \varphi\right)^{2}-\frac{\zeta^{2}}{64
                         \epsilon^{2n+1}}I_{2n+1/2}+\frac{\alpha}{2}\varphi^{2}
                         \right. \nonumber\\
                      &+&\left.\frac{\alpha^{2}}{64\pi^{2}}\left[\ln
                         \left(\frac{\alpha}{\mu_{R}^{2}}\right)-\frac{1}{2}
                         \right]-\frac{\left(\alpha-\mu_{R}^{2}\right)^{2}}{2
                         \lambda_{R}}+\frac{\mu_{R}^{4}}{2\lambda_{R}}
                         \right\} \;\;,
\end{eqnarray}

\begin{eqnarray}
\langle T_{yy} \rangle_{R}&=&l^{2}\langle T_{xx} \rangle_{R} \;\;,
\end{eqnarray}

\begin{eqnarray}
\langle T_{zz} \rangle_{R}&=&\left(\partial_{z}\varphi\right)^{2}+\langle 
T_{xx} \rangle_{R} \;\;.
\end{eqnarray}


\subsection{Renormalization of Off-Diagonal Components of 
Energy-Momentum Tensor}


First of all, we will analyze the off-diagonal elements of
$\langle T_{\mu \nu} \rangle$, when one of their index is the time $t$.
Considering the Gaussian state (\ref{gauss:pure}), these expectation
values are

\begin{eqnarray}
\langle T_{xt} \rangle&=&-\frac{{\hat{\pi}}^{2}}{2\epsilon}+A^{2}\beta^{x}
                         \left[-\frac{1}{2A^{2}}\left(\partial_{z}\varphi
                         \right)^{2}+\frac{\mu^{2}}{2}\varphi^{2}
                         +\frac{\lambda}{4!}\varphi^{4}\right] \nonumber\\
                      &-&\frac{2i\sigma}{\sqrt{\epsilon}}(G\partial_{x}\Omega)
                         ({\bf x},{\bf x},t)+A^{2}\beta^{x}\left\{\frac{2}
                         {\epsilon}(\Sigma\Omega\Sigma)({\bf x},{\bf x},t)
                         -\frac{1}{8}\Omega^{-1}({\bf x},{\bf x},t) \right. 
                         \nonumber\\
                      &+&\frac{1}{2A^{2}}\left[{\partial_{x}^{(\bf x)}}^{2}
                         +l^{-2}{\partial_{y}^{(\bf x)}}^{2}+{\partial_{z}^
                         {(\bf x)}}^{2}-2A^{2}\beta^{z}\partial_{x}^{(\bf x)}
                         \partial_{z}^{(\bf y)}\right]\Omega({\bf x},{\bf y},t)
                         \mid_{{\bf x}={\bf y}} \nonumber\\
                      &-&\left. \frac{1}{2}\left(\mu^{2}+\frac{\lambda}{2}
                         \varphi^{2}\right)\Omega({\bf x},{\bf x},t)-
                         \frac{\lambda}{8}\Omega({\bf x},{\bf x},t)
                         \Omega({\bf x},{\bf x},t)\right\} \nonumber\\       
                      &-&\beta^{x}{\partial_{x}^{(\bf x)}}^{2}\Omega({\bf x},
                         {\bf y},t)\mid_{{\bf x}={\bf y}} \;\;,
\label{Txt}
\end{eqnarray}

\begin{eqnarray}
\langle T_{yt} \rangle&=&-\frac{2i\sigma}{\sqrt{\epsilon}}(G\partial_{y}\Omega)
                         ({\bf x},{\bf x},t)-\left[\beta^{x}\partial_{y}^
                         {({\bf x})}\partial_{x}^{({\bf y})}+\beta^{z}\partial
                         _{y}^{({\bf x})}\partial_{z}^{({\bf y})}\right]\Omega
                         ({\bf x},{\bf y},t)\mid_{{\bf x}={\bf y}} \;\;,
\label{Tyt}
\end{eqnarray}

\begin{eqnarray}
\langle T_{zt} \rangle&=&\frac{\sigma}{\sqrt{\epsilon}}\hat{\pi}\partial_{z}
                         \varphi-\frac{{\hat{\pi}}^{2}}{2\epsilon}+A^{2}
                         \beta^{z}\left[-\frac{1}{2A^{2}}\left(\partial_{z}
                         \varphi\right)^{2}+\frac{\mu^{2}}{2}\varphi^{2}
                         +\frac{\lambda}{4!}\varphi^{4}\right] \nonumber\\
                      &-&\frac{2i\sigma}{\sqrt{\epsilon}}(G\partial_{z}\Omega)
                         ({\bf x},{\bf x},t)+A^{2}\beta^{z}\left\{\frac{2}
                         {\epsilon}(\Sigma\Omega\Sigma)({\bf x},
                         {\bf x},t)-\frac{1}{8}\Omega^{-1}({\bf x},{\bf x},t) 
                         \right. \nonumber\\
                      &+&\frac{1}{2A^{2}}\left[{\partial_{x}^{(\bf x)}}^{2}
                         +l^{-2}{\partial_{y}^{(\bf x)}}^{2}+{\partial_{z}^
                         {(\bf x)}}^{2}-2A^{2}\beta^{x}\partial_{z}^{(\bf x)}
                         \partial_{x}^{(\bf y)}\right]\Omega({\bf x},{\bf y},t)
                         \mid_{{\bf x}={\bf y}} \nonumber\\
                      &-&\left. \frac{1}{2}\left(\mu^{2}+\frac{\lambda}{2}
                         \varphi^{2}\right)\Omega({\bf x},{\bf x},t)
                         -\frac{\lambda}{8}\Omega({\bf x},{\bf x},t)
                         \Omega({\bf x},{\bf x},t)\right\} \nonumber\\       
                      &-&\beta^{z}{\partial_{z}^{(\bf x)}}^{2}\Omega({\bf x},
                         {\bf y},t)\mid_{{\bf x}={\bf y}} \;\;.
\label{Tzt}
\end{eqnarray}

\noindent
We use the expectation values (\ref{expvalue1}--\ref{expvalue2}) in 
order to obtain $\langle T_{ty} \rangle$

\begin{equation}
\langle T_{ty} \rangle= \langle T_{yt} \rangle= \frac{i\sigma}{\sqrt
{\epsilon}}\partial_{j}^{({\bf x})}\delta^{d}({\bf x}-{\bf
y})\mid_{{\bf x}} \;\;.
\label{Tyt:sim}
\end{equation}
Following the procedure used in the previous subsections, we can write 
(\ref{Txt}) and (\ref{Tzt}) in terms of renormalized coupling constants 
$\mu_{R}^{2}$ and $\lambda_{R}$, 

\begin{eqnarray}
\langle T_{xt} \rangle_{R}=\langle T_{tx} \rangle_{R}&=&-\frac{{\hat{\pi}}^{2}}
   {2\epsilon}+A^{2}\beta^{x}\left\{-\frac{1}{2A^{2}}\left(\partial_{z}
   \varphi\right)^{2}+\frac{\zeta^{2}}{64\epsilon^{2n+1}}I_{2n+1/2}
   -\frac{\alpha}{2}\varphi^{2}\right. \nonumber\\
&-&\left.\frac{\alpha^{2}}{64\pi^{2}}\left[\ln\left(\frac{\alpha}{\mu_{R}
   ^{2}}\right)-\frac{1}{2}\right]+\frac{{(\alpha-\mu_{R}^{2})}^{2}}{2\lambda
   _{R}}-\frac{\mu_{R}^{4}}{2\lambda_{R}}\right\} \;\;,
\end{eqnarray}

\begin{eqnarray}
\langle T_{zt} \rangle_{R}=\langle T_{tz} \rangle_{R}&=&\frac{\sigma}
   {\sqrt{\epsilon}}\hat{\pi}\partial_{z}\varphi-\frac{{\hat{\pi}}^{2}}
   {2\epsilon}+A^{2}\beta^{z}\left\{-\frac{1}{2A^{2}}\left(\partial_{z}
   \varphi\right)^{2}+\frac{\zeta^{2}}{64\epsilon^{2n+1}}I_{2n+1/2}
   -\frac{\alpha}{2}\varphi^{2}\right. \nonumber\\
&-&\left.\frac{\alpha^{2}}{64\pi^{2}}\left[\ln\left(\frac{\alpha}{\mu_{R}
   ^{2}}\right)-\frac{1}{2}\right]+\frac{{(\alpha-\mu_{R}^{2})}^{2}}{2\lambda
   _{R}}-\frac{\mu_{R}^{4}}{2\lambda_{R}}\right\} \;\;,
\end{eqnarray}
while $\langle T_{yt} \rangle$ give us an integral that vanishes in the
momentum space and, therefore,

\begin{equation}
\langle T_{yt} \rangle_{R}=0 \;\;.
\end{equation}
The spatial components of off-diagonal elements of $\langle T_{\mu
\nu} \rangle$ are easily calculated:

\begin{equation}
\langle T_{ij} \rangle=\langle T_{ji} \rangle=\partial_{i}^{({\bf x})}\partial_
{j}^{({\bf y})}\Omega({\bf x},{\bf y},t)\mid_{{\bf x}={\bf
y}} \;\;.
\label{Tij}
\end{equation}
It is easy to see that $\langle T_{ij} \rangle$ vanishes too since 
(\ref{Tij}) is also a null integral in the momentum space, like as 
(\ref{Tyt:sim}).

In the same way of ref.\cite{hug}, notice that the expectation of
value $\langle T_{\mu\nu} \rangle$ become finite in our Gaussian
approximation, once we take into account the mass and coupling
renormalization, and that $n>2$. This means that the infinities which
exist in the free scalar field model are canceled by the presence of
interactions, more specifically the term
$-g_{\mu\nu}\left(\lambda/8\right)\Omega\Omega^{\cite{ebo}}$.

In addition, almost all divergences that appear when $\lambda_{R}=0$
(free theory) vanishes in consequence of the structure equation of
motion for $\alpha$. The only infinite that survives is
$g_{\mu\nu}\mu_{R}^{4}/\lambda_{R}$, which is absorbed by a
redefinition of the cosmological constant.


\section{Conclusions and discussion}


The functional Schr\"odinger picture is a useful tool to the analysis
of initial value problems like the evolution of the early universe
during and/or after an inflationary period. In this work we establish
that a simple Gaussian variational approximation gives a consistent
picture of the time evolution of matter and gravitation when the space
is inhomogeneous in one dimension. In fact we were able to prove the
renormalizability not only of the equations of motion of the scalar
field, but also of the energy-momentum tensor. In order to obtain a
finite effective action we used the same renormalization prescription
that renders finite this quantity in the homogeneous
case$^{\cite{hug}}$. Moreover, the energy-momentum tensor is finite
once we substitute the bare couplings and masses by the renormalized
ones -- that is, we do not need to introduce further geometrical
counter-terms.  This is a remarkable result taking into account that
our approximation includes some non-linearities through the
self-consistent equations of motion.

In principle, the finite equations of motion for the scalar field
supplemented by the semi-classical Einstein equation, 

\begin{equation}
G_{\mu \nu}=-8\pi G_{N} \langle T_{\mu \nu} \rangle_{GR} \;\;,
\label{einst}
\end{equation}
where $\langle T_{\mu \nu} \rangle_{GR}$ is given by

\begin{equation}
\langle T_{\mu\nu} \rangle_{GR} \equiv  \langle T_{\mu\nu} \rangle_{R} -
g_{\mu\nu} \frac{\mu_R^4}{2\lambda_R} \; ,
\end{equation}
may be used to study dynamical questions about the universe for
inhomogeneous situations, such as stability of the space and the
conditions for inflation setting in. Nevertheless, we should keep in
mind that the Gaussian approximation in higher-dimensional field theory
suffers from well known shortcomings, analogous to the that appear in
the large-$N$ approximation. For instance, it does not describe
spontaneous symmetry breaking correctly in four dimensions.  In
addition to that the behavior of the kernels assumed in the simplified
{\em Ansatz} should describe well only the high frequency modes.
Therefore, we are confined to the study of the chaotic inflation
scenario within the framework developed here$^{\cite{num}}$.


\acknowledgments


I would like to thank Dr. O. J. P. \'Eboli for ideas and discussions to
this work and Dr. G. E. A. Matsas, for the kind hospitality of the
Instituto de F\'{\i}sica Te\'{o}rica, where a part of this work was
done. This work was supported by Funda\c{c}\~{a}o de Amparo \`{a}
Pesquisa do Estado de S\~{a}o Paulo (FAPESP), Funda\c{c}\~{a}o de
Amparo \`{a} Pesquisa do Estado do Rio Grande do Sul (FAPERGS) and by
Conselho Nacional de Desenvolvimento Cient\'{\i}fico e Tecnol\'{o}gico
(CNPq).




\begin{references}

\bibitem[*]{hcr} e-mail: hreis@charme.if.usp.br or hreis@ifi.unicamp.br 

\bibitem{abb} For a review see L.Abbott and S.-Y. Pi, {\it Inflationary 
              Cosmology} (World Scientific, Singapore, 1986); \\
              A. Linde, {\it Particle Physics and Inflationary
              Cosmology}, Harwood Academic Publishers, Chur,
              Switzerland, 1990.

\bibitem{alb} A. Albrecht, R. H. Brandenberger, and R. Matzner, 
              {\it Phys. Rev.} {\bf D32}, 1280 (1985).

\bibitem{suo} H. K.-Suonio, J. Centrella, R. Matzner, and J. R. Wilson, 
              {\it Phys. Rev.} {\bf D35}, 435 (1987).

\bibitem{ebo} O. \'Eboli, S.-Y. Pi, and M. Samiullah, {\it Ann. Phys.} 
              {\bf 193}, 102 (1989); \\ M. Samiullah, O. J. P. \'Eboli,
              and S.-Y. Pi, {\it Phys. Rev.}  {\bf D44}, 2335 (1991).

\bibitem{hug} H. C. Reis and O. J. P. \'Eboli, {\it Int. J. Mod. Phys.}
              {\bf A11}, 3957 (1996).

\bibitem{jac} R. Jackiw, {\it Int J. Quantum Chem.} {\bf 17}, 41 (1980).

\bibitem{ker} R. Jackiw and A. Kerman, {\it Phys. Let.} {\bf A71}, 158 (1979).

\bibitem{bal} R. Balian and M. V\'{e}n\'{e}roni, {\it Phys. Rev. Lett.}
              {\bf 47}, 1353 (1981); {\bf 47}, 1765 (1981); \\
              O. \'Eboli, R. Jackiw, and S.-Y. Pi, {\it Phys. Rev.}
              {\bf D37}, 3557 (1988).

\bibitem{sch} For a review see R. Jackiw and S.-Y. Pi contributions to
              {\it Field Theory and Particle Physics}, V Jorge Andr\'e 
              Swieca Summer School, Campos do Jord\~{a}o, Brazil, 1989,
              edited by O. J. P. \'Eboli, M. Gomes, and A. Santoro (World 
              Scientific, Singapore, 1990).

\bibitem{sym} K. Symanzik, {\it Nucl. Phys.} {\bf B190}, 1 (1983).

\bibitem{sam} S.-Y. Pi and M. Samiullah, {\it Phys. Rev.} {\bf D36}, 
             3128 (1987).

\bibitem{dir} P. A. M. Dirac, {\it Proc. Cambridge Philos. Soc.} {\bf
              26}, 376 (1930).

\bibitem{col} G.\ 't Hooft and M.\ Veltman, Nucl.\ Phys.\ {\bf B44}, 189
              (1972); C.\ G.\ Bollini and J.\ J.\ Giambiagi, Nuovo Cim.\ {\bf
              12B}, 2 (1972); J. Collins, {\it Renormalization} (Cambridge
              University Press, 1984).

\bibitem{num} H. C. Reis, work in progress.

\end{references}
\end{document}